\documentclass[sigconf]{acmart}
\AtBeginDocument{%
  \providecommand\BibTeX{{%
    \normalfont B\kern-0.5em{\scshape i\kern-0.25em b}\kern-0.8em\TeX}}}
\usepackage{booktabs}
\usepackage{multirow}
\usepackage{xcolor}
\usepackage{listings}
\setcopyright{acmcopyright}
\copyrightyear{2027}
\acmYear{2027}
\acmDOI{XXXXXXX.XXXXXXX}
\acmConference[WWW '27]{The ACM Web Conference 2027}{May 10--14, 2027}{Dublin, Ireland}
\acmBooktitle{Proceedings of the ACM Web Conference 2027 (WWW '27), May 10--14, 2027, Dublin, Ireland}
\acmYear{2027}

\begin{document}

\title{Fear Doesn't Go Viral, It Lingers: Spread, Affect, and Discourse of Food-Fear Content on TikTok}

\author{Hunjun Shin}
\orcid{0009-0003-5936-4933}

\affiliation{%
  \institution{Northeastern University}
  \city{Boston}
  \state{Massachusetts}
  \country{USA}
}
\email{shin.hu@northeastern.edu}

\author{Hongpeng Fu}
\affiliation{%
  \institution{Northeastern University}
  \city{Seattle}
  \state{Washington}
  \country{USA}
}
\email{fu.hongp@northeastern.edu}

\author{Divya Chaudhary}
\orcid{0000-0002-8877-2034}
\affiliation{%
  \institution{Northeastern University}
  \city{Seattle}
  \state{Washington}
  \country{USA}
}
\email{d.chaudhary@northeastern.edu}

\author{Mohit Singhal}
\orcid{0000-0002-7423-9116}
\affiliation{%
  \institution{Northeastern University}
  \city{Boston}
  \state{Massachusetts}
  \country{USA}
}
\email{m.singhal@northeastern.edu}

\renewcommand{\shortauthors}{Hunjun Shin et al.}

\begin{abstract}
TikTok has become a primary channel for nutrition information, and alongside evidence-based content it hosts a growing stream of \emph{fear-based} wellness messaging that frames ordinary foods as ``toxic,'' ``poison,'' or things to ``never eat.'' Prior work has extensively studied political and public-health misinformation and audited the accuracy of health videos, but to the best of our knowledge, no work has examined how everyday food-fear content spreads and how audiences respond to it. To investigate this, we collected 97,234 TikTok food and nutrition videos, labeled them \textit{Fear} or \textit{Non-Fear} with a zero-shot multimodal classifier built on Gemini~3 Flash that is validated against human consensus (Cohen's $\kappa=0.81$), and analyzed more than 1.1M comments and 272K parent-reply pairs. Contrary to the intuition that fear ``goes viral,'' fear videos reach fewer people than non-fear videos. Their distinctive signature lies in dynamics: they are shared more per view (1.7--2.1$\times$ after controls) and their discussion persists longer (comment half-life 1.62$\times$ after controls). These shifts are modest in size but hold after controlling for reach. Comments on fear videos are more negative, angrier, and more disgusted, yet negativity is \emph{not} more self-reinforcing: negative comments draw negative replies at the same rate as in non-fear threads, while positive comments are far less likely to stay positive. Fear threads are contested rather than one-sided, with both more corrective and more alarm-amplifying replies. 
Fear discourse concentrates in frames about industrial ingredients, contaminants, and institutional distrust, and its repertoire of frames widens over 2020--2026. Our results argue for persistence-based rather than virality-based signals when platforms seek harmful health content. Because nearly everyone makes food choices daily, understanding how fear about food takes hold online matters for public health.

\end{abstract}



\maketitle

\section{Introduction}
Misinformation research has concentrated on high-stakes, event-driven domains such as false news during elections \citep{allcott2017social,grinberg2019fake} and health misinformation during crises such as the COVID-19 pandemic and vaccine debates \citep{cinelli2020covid,broniatowski2018weaponized,loomba2021measuring}. In these settings, misleading content is often studied as something that spreads fast and wide, and false stories can indeed outpace true ones \citep{vosoughi2018spread}. A quieter and more persistent genre has received far less attention: everyday fear-mongering about food. Short videos tell viewers that ordinary foods, ingredients, and brands are ``toxic,'' ``poison,'' or things to ``never eat,'' usually without proportionate evidence. Food choices are made daily by everyone, so such messages reach a far broader and less politicized audience than election or vaccine content, and they draw on well-documented biases in how people judge chemical and dietary risk \citep{siegrist2019chemophobia}. TikTok is now a primary source of nutrition information for many users, and its nutrition content is often of low quality, with less accurate posts drawing more engagement \citep{zeng2025whatieat}.

The genre has a recognizable style. Creators open with a threat and a promise of secret knowledge: they will tell you what is really in your food because nobody else will; an ingredient in a popular cereal is the same chemical used in yoga mats; a common snack is banned in other countries but sold freely at home; seed oils are industrial oils processed with a solvent from gasoline; parents should throw a breakfast staple away today. Many videos list several foods to avoid and ask viewers to watch until the end for the worst one. In our corpus, 62\% of fear videos use at least one such hook in their description or speech, against 37\% of non-fear videos, and some hooks are almost exclusive to fear content: claims that a food causes cancer or is ``killing you'' are 20 times more common, ``banned in other countries'' claims are 45 times more common, and appeals to hidden truth are 10 times more common (Table~\ref{tab:hooks}). The messages rarely come with a dose, a source, or a safer action, and they ask viewers to distrust both the food on their shelves and the agencies that regulate it.

\begin{figure}[t]
\centering
\includegraphics[width=0.85\columnwidth]{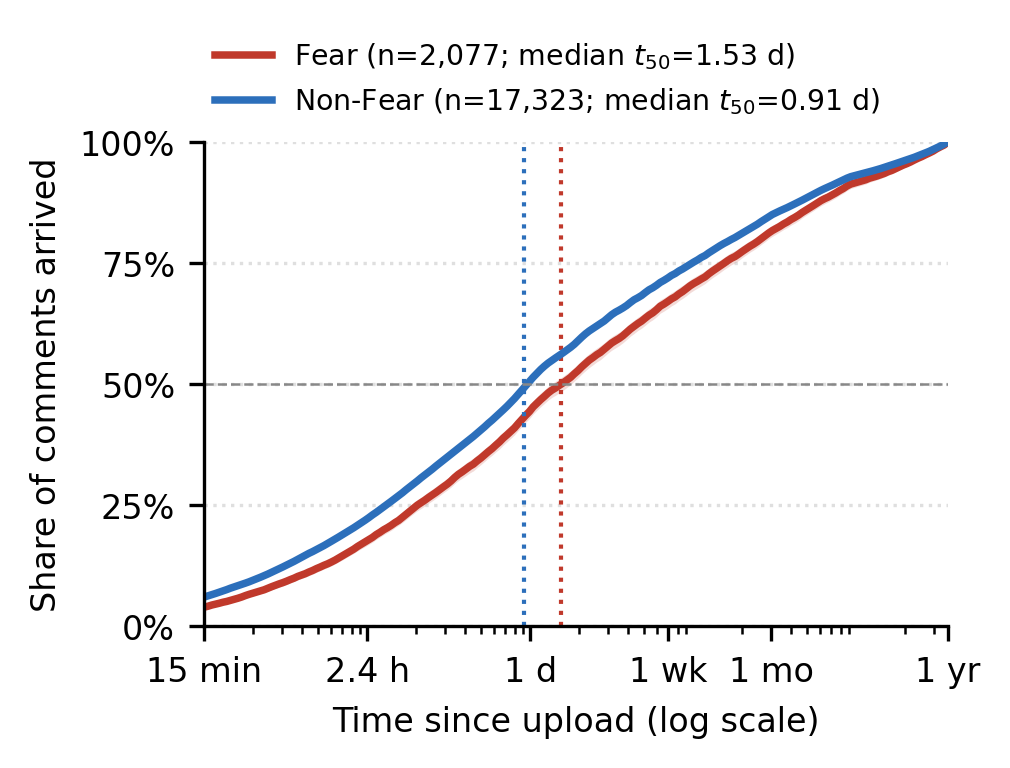}
\caption{Fear doesn't spike; it lingers. Average share of a video's comments that have arrived by each time since upload, with 95\% video-level bootstrap bands. One day after upload, 45\% of a fear video's comments have arrived versus 51\% for non-fear videos. Dotted lines mark the median comment half-life $t_{50}$. 
}
\label{fig:arrival}
\end{figure}

This is significant because the harms of food fear are scattered but real. Health misinformation on social media harms public health \citep{chou2018addressing,suarez2021prevalence,kirkpatrick2024tiktok}, and exaggerated food risks can push people away from affordable, nutritious foods, feed anxiety and restrictive eating, and erode trust in food-safety institutions, all without a single false claim ever trending. Platforms, meanwhile, have built their responses to health misinformation around high-salience events such as COVID-19, where warning labels and takedowns target content that is spreading quickly \citep{singhal2023sok,ling2023learn}. Prior evaluations of these tools study COVID-19 warning labels on TikTok \citep{ling2023learn}, soft moderation of election claims on Twitter \citep{zannettou2021won}, community notes on X \citep{drolsbach2023diffusion,chuai2024community}, and health professionals who debunk claims on TikTok \citep{sharevski2024debunk,bhargava2023effective}. They examine interventions on content selected because it is false, viral, or tied to a news event. We instead characterize the organic dynamics of a content category that is rarely false outright and rarely viral, which is the setting these interventions would have to work in. 
To the best of our knowledge, no prior work has examined how food-fear content behaves once it is posted. A natural intuition, carried over from election and crisis misinformation, is that fear content succeeds because it ``goes viral.'' In the food domain, we find that this intuition is misleading. Fear videos reach \emph{fewer} people than non-fear videos; their distinctive signature lies in how attention unfolds after posting: more sharing per view, discussion that lasts longer (Figure~\ref{fig:arrival}), and comment sections with a different emotional and conversational character. We organize the analysis around three research questions:

\begin{itemize}
\item \textbf{RQ1 (Spread).} Do fear-based food videos spread and sustain
discussion differently from non-fear videos?
\item \textbf{RQ2 (Affect).} What affective and conversational environment do
they create in comments and replies?
\item \textbf{RQ3 (Discourse).} What is fear discourse about, and how does it
change over time?
\end{itemize}

To answer these questions we collect food and nutrition videos through the TikTok Research API~\citep{tiktokapi} using fear-oriented and neutral control keywords, draw a stratified sample of 134{,}365 videos, and label 97{,}234 of them as \textit{Fear} or \textit{Non-Fear} with \texttt{gemini-\allowbreak 3-\allowbreak flash-\allowbreak preview} \citep{google2026gemini3flash}, which watches each video and labels it by following a written codebook; we validate its labels against two human annotators. For RQ1, we compare per-view engagement and the timing of comment arrival, using regressions that control for reach, video length, upload year, and keyword. For RQ2, we label the sentiment and emotions of comments and model how each reply's sentiment depends on its parent comment, and we label replies for corrective and amplifying stance. For RQ3, we fit topic models to video transcripts and comments and track fear-dominated topics over time.

Our key findings are (1) fear videos get 62\% fewer views but are shared 1.7--2.1$\times$ more per view and keep their discussion going 1.6$\times$ longer, (2) their comment sections are angrier and less able to sustain positivity, but are not echo chambers, since negativity is reinforced at the same rate as in non-fear threads and threads contain more corrective replies, and (3) fear discourse targets industrial ingredients, contaminants, and institutions, with a repertoire of frames that widens over time. We make the following contributions: 

\begin{enumerate}
\item \textbf{Persistence over reach:} Early evidence that the influence of food-fear content lies in how long attention lasts rather than how far it spreads, robust to controls, video age, comment coverage, and classifier errors.
\item \textbf{Asymmetric damping, not echo chambers:} A reply-level account of fear threads in which positivity fails to persist while negativity is reinforced no more than in non-fear threads, and corrective and amplifying voices coexist.
\item \textbf{A widening repertoire:} A map of what food fear is about and how its frames multiply over time, with implications for detection, soft moderation, and community-based correction.
\end{enumerate}
Together, these results show that the standard picture of harmful content as something that spikes and spreads does not fit one of the most common forms of health misinformation people encounter. Food fear reaches viewers who return to it, share it, and argue about it for weeks, so detection, moderation, and correction strategies built for viral misinformation will tend to miss it. Understanding its slower dynamics is a first step toward interventions that fit how this content actually behaves.

\section{Related Work}
\textbf{Health misinformation on social and video platforms.}
Prior works have found health misinformation across platforms and topics, from vaccines and infectious disease to diet and nutrition \citep{wang2019systematic,suarez2021prevalence}. Video platforms are a particular concern: over a quarter of the most-viewed COVID-19 videos on YouTube contained misleading information \citep{li2020youtube}, and audits show that search and recommendation can steer users toward misinformation once they have watched it \citep{hussein2020measuring}. On TikTok, diet and nutrition content is dominated by weight-normative messaging and diet culture \citep{minadeo2022weight,munro2024diet}; much of it is of low quality, and less accurate posts attract \emph{more} engagement \citep{zeng2025whatieat}. Food fears also draw on documented biases in lay risk perception, such as the tendency to judge synthetic chemicals as more dangerous than natural ones \citep{siegrist2019chemophobia}. Most of this work audits the accuracy of videos about a specific condition. We instead characterize how an entire fear-oriented content category behaves: how it spreads, what conversations it produces, and how its frames change.

\textbf{Attention dynamics and diffusion on the Web.} A long line of Web research models how attention to online content rises and decays~\citep{wu2007novelty}. 
YouTube view series fall into dynamic classes with power-law relaxation after a burst \citep{crane2008robust}, Twitter hashtags show distinct classes of activity around their peak \citep{lehmann2012dynamical}, and early popularity and cascade structure predict later reach \citep{szabo2010predicting,cheng2014cascades}. Emotional and moral content spreads further \citep{berger2012viral,brady2017emotion}, negativity increases consumption \citep{robertson2023negativity}, out-group animosity drives engagement \citep{rathje2021outgroup}, and false news can diffuse faster than true news \citep{vosoughi2018spread}. These studies mostly ask how far and how fast content travels. We add the shape of the attention curve as an outcome.

\textbf{Affect, contagion, and echo chambers in comment threads.}
Exposure to emotional posts shifts the emotion of users' own posts \citep{kramer2014experimental,ferrara2015measuring}, and social feedback reinforces moral outrage over time \citep{brady2021social}. Misinformation tends to circulate within homogeneous communities \citep{delvicario2016spreading}, and the strength of such echo chambers differs across platforms \citep{cinelli2021echo}. These accounts predict that fear threads should amplify negativity. We test that prediction at the level of parent--reply pairs and find instead that negativity is reinforced at the same rate as in non-fear threads while positivity fails to persist.

\textbf{Correction and counter-speech.} Users and platforms can correct misinformation in place. Corrections delivered through related content and by expert sources reduce health misperceptions \citep{bode2015related,vraga2017using}, ordinary users write a large share of the replies that counter misinformation \citep{micallef2020role}, and counter-speech is common in hateful comment threads \citep{mathew2019thou}. Whether such voices are present in organic fear content has not been measured. We find that fear threads contain more corrective replies, and more alarm-amplifying ones, than other food threads.

\textbf{Fear appeals.} Health-communication theory, notably the Extended Parallel Process Model (EPPM) \citep{witte1992putting} and meta-analyses \citep{witte2000meta,tannenbaum2015appealing}, holds that fear appeals change behavior when paired with efficacy and otherwise trigger defensive ``fear control'' (reactance, avoidance, distrust). We study fear appeals as they occur organically in a social feed and read audience emotions through this lens.

\section{Methodology}
\textbf{Video collection.} We designed 97 fear-oriented search queries spanning ten kinds of fear framing found in food content (toxicity and poison, processed food, health effects such as cancer or inflammation, fear-inducing phrases, authority warnings such as ``doctors never eat,'' hidden secrets and industry lies, avoid or never-eat appeals, additives and chemicals, bans and recalls, and specific diets), and 49 neutral control queries spanning four kinds of everyday nutrition content (healthy eating and recipes, nutrition education, specific nutrients, and meal prep and cooking). We queried the TikTok Research API \citep{tiktokapi} with each one; 66 fear and 37 control queries returned videos (e.g., ``toxic food,'' ``foods to avoid,'' ``never eat this''; ``healthy recipes,'' ``meal prep ideas''), retrieving about 677K videos with their metadata and engagement counts (views, likes, shares, comments). Because the API matches query terms anywhere in a video's description, results include unrelated content, which is why we label every sampled video rather than relying on its keyword. 
To keep keyword sets comparable over time, we retained keywords with at least 1{,}000 videos and at least 48 months of coverage, leaving 17 fear and 22 control keywords (full list in Appendix~\ref{app:keywords}). From these, we drew a stratified sample by keyword and upload year with square-root proportional allocation, capping each creator at two videos so that no single account dominates, which yielded 134{,}365 candidate videos. We collected video metadata through the Research API between February 14 and March 2, 2026, covering videos uploaded from December 2018 to February 2026; comments were collected on April 6--8, 2026.

\textbf{Video files, transcripts, and comments.} We downloaded the public video files of sampled videos with an automated browser and processed 98{,}042 of them; 97{,}234 received a valid Fear/Non-Fear label. We transcribed the audio with \texttt{gemini-\allowbreak 2.0-\allowbreak flash} \citep{gemini2024,comanici2025gemini}, following prior work that transcribed TikTok videos with the same model \citep{balloli2026reliance}; Gemini models match or exceed dedicated speech recognizers such as Whisper on long-form video transcription \citep{gemini2024,radford2023whisper}. Speech was found in 81{,}595 videos; the 76{,}713 English transcripts are used for topic modeling. Comments and their replies were collected from public TikTok pages in April 2026 (April 6--8 for the comments analyzed here), up to about 500 per video in TikTok's default display order, and deduplicated by comment identifier. The collection returned a median of 67\% of each video's reported comment count. Appendix Table~\ref{tab:data} summarizes the analysis samples. We did not collect creator follower counts, so creator size is not observed.

\textbf{Definition.} We label a video \emph{Fear} when it uses fear-inducing language, alarmist claims, urgency, or sensational framing to create unnecessary fear about food or nutrition. The codebook lists ten verbal and visual indicators (e.g., alarmist terms such as ``toxic'' or ``poison,'' urgency, absolutist claims, hidden-truth or conspiracy framing, warning imagery, rejection imagery); a video with two or more indicators is \textit{Fear}, and one with a single indicator is \textit{Fear} only if its overall tone is alarming. We label those that do not meet the criteria as \emph{Non-Fear}, including responsibly presented recalls and allergen alerts, balanced health information, and satire or debunking of fear claims. The full codebook is in Appendix~\ref{app:codebook}.

\textbf{Human annotation.} Two annotators independently watched and labeled a stratified sample of 1{,}000 videos (500 per keyword group); both labeled 986. The other 14 videos had playback errors (11 of them also failed automated classification), so our gold set consists of the 986 double-labeled videos. 
Agreement was 90.5\% with Cohen's $\kappa=0.58$. Because Fear is the minority class, $\kappa$ is much lower than raw agreement. Agreement on such subjective, intent-laden constructs is often lower still, as reported in prior work with Krippendorff's $\alpha$ of 0.30--0.51 for fear speech \citep{saha2023rise} and 0.18--0.29 for hate speech \citep{ross2016measuring}.

\textbf{LLM classification.} We classified every video with a zero-shot multimodal large language model (\texttt{gemini-\allowbreak 3-\allowbreak flash-\allowbreak preview}, the Gemini 3 Flash model \citep{google2026gemini3flash}, temperature 0.1) that receives the video itself (audio and frames) and the codebook (Appendix~\ref{app:prompt}).
We used a low temperature to keep labels nearly deterministic; we did not tune this value. 
The prompt also passes the search keyword and its group, which could nudge labels toward the keyword set. To test this, we re-classified the human-labeled videos twice, with the production prompt and with the keyword information removed ($n=770$ with valid labels under both). The two prompts agree on 98.4\% of videos ($\kappa=0.94$), label the same share as Fear (16.2\%), and match human labels equally well on these 770 videos (Fear precision 0.72 vs.\ 0.71, recall 0.92 vs.\ 0.91), so the keyword information does not drive the labels. Of the double-labeled videos, 819 are in the classified corpus, 817 of which have an adjudicated label. On the 734 where both annotators agreed, the classifier matches the human label 96.3\% of the time (Cohen's $\kappa=0.81$, Fear-F1 0.83), with very high recall for Fear (0.99) and moderate precision (0.72). Against the adjudicated final label on all double-labeled videos, which includes the harder cases where annotators disagreed, Fear precision is 0.64. The classifier is therefore a sensitive detector that over-labels \textit{Fear} because misclassified Non-Fear videos dilute the Fear group; this error biases group differences toward zero. We test this directly with label-noise robustness checks, as shown in Section~\ref{sec:robust}.

\begin{table}[t]
\centering
\footnotesize
\caption{Share of videos using common fear hooks, matched with regular
expressions. All differences are Holm-adjusted $p<10^{-70}$. ``Toxic'' and ``stop
eating'' overlap with search keywords, so they also appear in many non-fear videos.
}
\begin{tabular}{lrrr}
\toprule
Hook (description or speech) & Fear & Non-Fear & Ratio \\
\midrule
Any hook below & 61.8\% & 37.4\% & 1.7$\times$ \\
\midrule
Toxic / poison & 31.8\% & 16.8\% & 1.9$\times$ \\
Never / stop eating & 27.2\% & 20.2\% & 1.3$\times$ \\
Causes cancer / killing you & 6.3\% & 0.3\% & 19.5$\times$ \\
Hidden truth (``they won't tell you'') & 4.6\% & 0.5\% & 10.1$\times$ \\
Industry or regulator distrust & 3.1\% & 0.3\% & 12.5$\times$ \\
Urgency or shock (``watch until the end'') & 2.9\% & 0.4\% & 7.3$\times$ \\
Banned in other countries & 2.2\% & 0.05\% & 45.1$\times$ \\
\bottomrule
\end{tabular}

\label{tab:hooks}
\end{table}

\section{Analysis Methods}
\textbf{Spread metrics.} Per-view \emph{share rate} is shares divided by views. For videos with at least five comments in a 0--365 day window, comment arrival percentiles $t_{25}$, $t_{50}$, $t_{75}$ are days since upload; $t_{50}$ is the discussion half-life. We also fit a per-video truncated power-law to comment arrival times on [15 min, 365 d] by maximum likelihood; a smaller exponent $\alpha$ means slower decay.

\textbf{Affect.} We label comments and replies as positive, neutral, or negative using \texttt{twitter-\allowbreak roberta-\allowbreak base-\allowbreak sentiment-\allowbreak latest} \citep{loureiro2022timelms,camacho2022tweetnlp}, a RoBERTa model trained on tweets and fine-tuned on TweetEval \citep{barbieri2020tweeteval}. To check it, one author labeled 208 comments drawn at random within each model label and keyword group. Agreement is moderate (accuracy 68\%, macro-F1 0.69, $\kappa=0.53$): the model catches 91\% of comments the annotator judged negative but often reads neutral comments as positive or negative. Because the sample is stratified by model label, we can reweight it to estimate the true share of negative comments. The model overstates negativity in both groups, but the gap survives: 24.4\% of comments under fear videos are negative versus 12.9\% under non-fear videos, a difference of 11.5 pp (95\% bootstrap CI 1.9--19.6), close to the 11.6 pp the model gives.  
For emotions we label a stratified sample of 150{,}000 English comments with the 11-label \texttt{twitter-\allowbreak roberta-\allowbreak base-\allowbreak emotion-\allowbreak multilabel-\allowbreak latest} model~\citep{camacho2022tweetnlp}, trained on the SemEval-2018 affect data \citep{mohammad2018semeval} (threshold 0.5).

\textbf{Reply direction and stance.} For each reply, we take the sentiment of the reply and its parent comment and model the probability that a reply is negative (or positive) as a function of Fear, parent sentiment, and their interaction, using GEE logistic models with robust standard errors clustered by video. Clustering matters because replies in the same thread are not independent, and the interaction directly tests whether the Fear effect differs by parent sentiment. For stance, we label 3{,}000 replies (1{,}500 per group, stratified by parent sentiment) with \texttt{gemini-\allowbreak 3-\allowbreak flash-\allowbreak preview} as agreeing or disagreeing with the parent, \emph{corrective} (challenging or correcting a health claim), and \emph{amplifying} (escalating alarm). One author manually annotated 100 replies, sampled to oversample those Gemini labeled corrective or amplifying so that these rare labels could be checked; Gemini agreed with the author on 88\% of stance labels ($\kappa=0.81$), 90\% of corrective labels ($\kappa=0.78$; precision 0.81), and 93\% of amplifying labels ($\kappa=0.84$; precision 0.82). 

\textbf{Topics.} We model transcripts and comments separately with BERTopic \citep{grootendorst2022bertopic} using \texttt{all-\allowbreak MiniLM-\allowbreak L6-\allowbreak v2} embeddings, filtering numeric tokens. We disable automatic topic merging and reassign HDBSCAN outliers to their nearest topic, which avoids a dominant catch-all topic (the comment model has 1{,}825 topics, the largest holding 1.2\% of comments). We take topics whose Fear share is at least twice the corpus baseline, read their representative documents, group them into categories, and exclude noisy, off-domain, or non-English clusters (Appendix~\ref{app:topics}).

\textbf{Statistics.} In our dataset, engagement counts, rates, and arrival times are heavily skewed, so we compare them with Mann--Whitney U tests, and compare proportions with two-proportion $z$ tests. Given we have tens of thousands of videos, almost any difference is significant, so we report effect sizes (Cliff's $\delta$ and Cohen's $h$) and interpret magnitude rather than $p$-values alone. We correct the 16 headline tests for multiple comparisons with the Holm--Bonferroni procedure~\citep{holm1979simple}. Because fear and non-fear videos differ in reach, length, upload year, and search keyword, raw gaps could reflect composition rather than content. We therefore fit regressions with log views, video-duration bins, upload-year and keyword-group fixed effects, and standard errors clustered by creator; $t_{50}$ and negativity models also control for log comment count. A stricter within-creator specification adds creator fixed effects for creators who posted both Fear and Non-Fear videos.

\section{Results}

\subsection{Fear Reaches Fewer People but Lingers Longer (RQ1)}
We found that \textit{Fear} content does not have a higher reach than \textit{Non-Fear}. Fear videos have lower median views (1{,}033 vs.\ 2{,}741) and likes (36 vs.\ 84) than non-fear videos (Table~\ref{tab:c1}; Cliff's $\delta=-0.21$ and $-0.22$). Their per-view like rate is also slightly lower. What distinguishes them is sharing and persistence. Fear videos are shared at 2.32$\times$ the median per-view rate of non-fear videos, 
and their comments arrive later at every percentile: the discussion half-life is 1.82 days versus 1.03 days. Comment arrival also decays more slowly with a median power-law exponent of 0.958 vs.\ 1.007.

These differences are not explained by reach or composition. With controls for log views, duration, upload year, and keyword group, the Fear share-rate ratio is 1.71$\times$ (log OLS) or 2.05$\times$ (Poisson pseudo-likelihood with a views offset), the half-life ratio is 1.62$\times$, and the decay-exponent difference is $-0.063$ [$-0.080$, $-0.047$], all $p<10^{-12}$. The effects hold within the fear and control keyword sets separately (the PPML share-rate estimate is borderline in the control set). They are modest in size: Cliff's $\delta$ is 0.16 for share rate and 0.13 for $t_{50}$, hence, the two distributions overlap substantially.

Typical high-share fear videos pair a threat with a hook. One video, shared 5{,}600 times, opens with ``You're being poisoned. Four popular foods you must avoid. Watch until the end because the last one will shock you.'' Another, shared 36{,}891 times from 406K views, reports a nationwide recall of food that ``could unalive you.'' Long-lived threads form around authority-styled warnings: in a video seen 693K times, a clinician says ``Takis? Stop eating them. Stop letting your kids eat them\ldots{} They're literally garbage,'' and half of its comments arrived more than 234 days after upload.

\begin{table}[t]
\centering
\small
\caption{Reach, engagement efficiency, and discussion persistence (medians).
Raw differences are Mann--Whitney U tests, Holm-adjusted $p<10^{-16}$ except the
comment rate. Adjusted = Fear coefficient (ratio, or difference for $\alpha$)
from regressions with log views, duration, year and keyword-group fixed effects,
creator-clustered SEs; all $p<10^{-12}$.}
\begin{tabular}{lrrrr}
\toprule
Metric & Fear & Non-Fear & Raw ratio & Adjusted \\
\midrule
Views (median)      & 1{,}033 & 2{,}741 & 0.38$\times$ & --- \\
Likes (median)      & 36      & 84      & 0.43$\times$ & --- \\
Share rate          & 0.00142 & 0.00061 & 2.32$\times$ & \textbf{1.71$\times$} \\
Like rate           & 0.0290 & 0.0325 & 0.89$\times$ & --- \\
Comment rate        & 0.00065 & 0.00057 & n.s. & --- \\
\midrule
Comment $t_{25}$ (d) & 0.572 & 0.285 & 2.01$\times$ & --- \\
Comment $t_{50}$ (d) & 1.82  & 1.03  & 1.76$\times$ & \textbf{1.62$\times$} \\
Comment $t_{75}$ (d) & 6.89  & 4.37  & 1.58$\times$ & --- \\
Decay exponent $\alpha$ & 0.958 & 1.007 & $-0.049$ & \textbf{$-0.063$} \\
\bottomrule
\end{tabular}

\label{tab:c1}
\end{table}

\subsection{Fear Threads Are Angrier but Not Echo Chambers (RQ2)}

\begin{figure}[t]
\centering
\includegraphics[width=0.9\columnwidth]{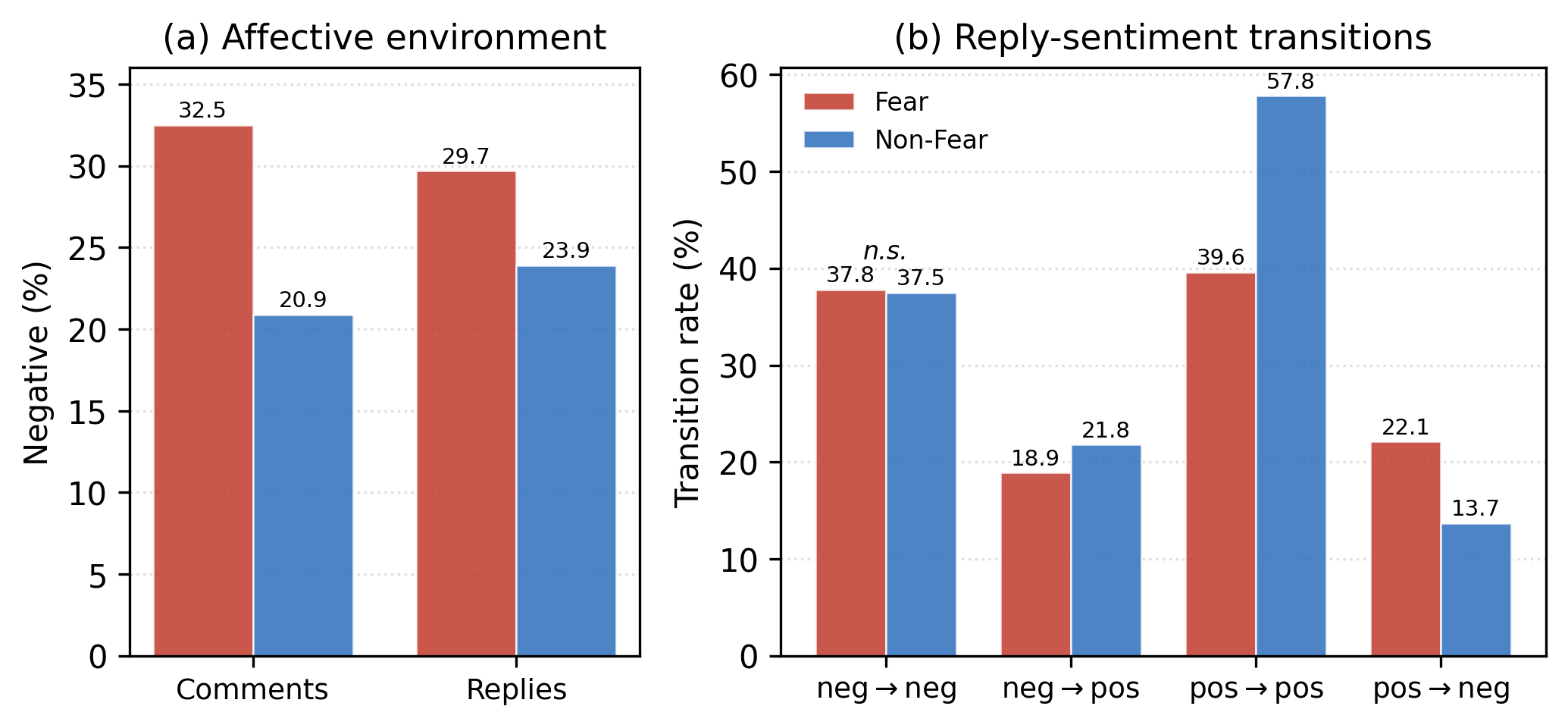}
\caption{The affective environment of fear content. (a) Negativity is higher for
fear at both the comment and reply level. (b) Reply-sentiment transitions:
negative$\rightarrow$negative reinforcement is equal across groups, while fear
threads retain far less positive continuity and convert positive parents to
negative replies more often.}
\label{fig:rq2}
\end{figure}

\textbf{More negative, angrier, more disgusted.} Fear content sits in a more negative comment environment (Figure~\ref{fig:rq2}a, Table~\ref{tab:c2}): 32.5\% of its comments are negative versus 20.9\% (Cohen's $h=0.27$). Per video, the gap is the largest effect in our study (Cliff's $\delta=0.39$) and survives controls (+9.6 pp). The emotion labels show the type of negativity. Fear videos do not make commenters much more afraid (4.6\% vs.\ 3.4\%, $h=0.06$). Instead, their comments carry more disgust (26.0\% vs.\ 16.9\%) and anger (21.9\% vs.\ 14.0\%), each with an adjusted odds ratio of 1.49, and less joy (31.7\% vs.\ 49.6\%), love, and optimism. Highly liked negative comments under fear videos voice exhaustion and blame rather than fright: ``It's not just the food. It's literally everything\ldots{} It's all designed to keep us busy and sick'' (37{,}831 likes); ``I just don't understand why all those products are not banned here in the USA, FDA needs to do better!'' (17{,}704 likes); and simply ``dont even know what to eat anymore'' (23{,}898 likes). Two of the three direct their anger at regulators and the food system rather than at the creator, foreshadowing the institutional distrust we find in RQ3. 

\begin{table}[t]
\centering
\small
\caption{Negativity and emotions in comments (per-video values are medians). All
differences are Holm-adjusted $p<10^{-30}$.}
\begin{tabular}{lrrr}
\toprule
Measure & Fear & Non-Fear & Effect \\
\midrule
Comment negative (aggregate) & 32.5\% & 20.9\% & $h=0.27$ \\
Comment negative (per video) & 0.286 & 0.133 & $\delta=0.39$ \\
Reply negative (aggregate) & 29.7\% & 23.9\% & $h=0.13$ \\
\midrule
Emotion: disgust & 26.0\% & 16.9\% & $h=0.22$ \\
Emotion: anger & 21.9\% & 14.0\% & $h=0.21$ \\
Emotion: fear & 4.6\% & 3.4\% & $h=0.06$ \\
Emotion: joy & 31.7\% & 49.6\% & $h=-0.37$ \\
Emotion: love & 6.5\% & 17.1\% & $h=-0.34$ \\
\bottomrule
\end{tabular}

\label{tab:c2}
\end{table}

\textbf{Asymmetric damping of positivity.} A plausible interpretation of the negativity gap is that fear threads are echo chambers in which negativity begets negativity. Our reply-direction models of 271{,}960 parent--reply pairs in 19{,}543 videos do not support that mechanism (Table~\ref{tab:c3}). A negative parent draws a negative reply at the same rate in fear and non-fear threads (37.8\% vs.\ 37.5\%, OR 1.01, n.s.). What differs is positivity: in fear threads, a positive parent is far less likely to draw a positive reply (39.6\% vs.\ 57.8\%, OR 0.48) and more likely to draw a negative one (22.1\% vs.\ 13.7\%, OR 1.78; Figure~\ref{fig:rq2}b). The Fear effect differs across parent sentiments (Wald $p<10^{-13}$), so we describe fear threads as damping positivity rather than reinforcing negativity. In one fear thread, the comment ``I am a colon cancer stage 3 survivor. Over 18yrs\ldots{} Thank Jesus \ensuremath{\heartsuit}'' (15{,}428 likes) drew the reply ``Don't it's literally mystery meat''; under fear videos, positive comments like this rarely receive a positive reply.

\begin{table}[t]
\centering
\small
\caption{Reply-direction transitions. Odds ratios for Fear vs.\ Non-Fear within
each parent sentiment from GEE logistic models clustered by video
($N=271{,}960$ pairs, 19{,}543 videos).}
\begin{tabular}{lrrl}
\toprule
Parent $\rightarrow$ reply & Fear & Non-Fear & OR [95\% CI] \\
\midrule
Negative $\rightarrow$ negative & 37.8\% & 37.5\% & 1.01 [0.93, 1.10] \\
Negative $\rightarrow$ positive & 18.9\% & 21.8\% & 0.84 [0.77, 0.91] \\
Positive $\rightarrow$ positive & 39.6\% & 57.8\% & 0.48 [0.41, 0.55] \\
Positive $\rightarrow$ negative & 22.1\% & 13.7\% & 1.78 [1.56, 2.03] \\
\bottomrule
\end{tabular}

\label{tab:c3}
\end{table}

\textbf{Contested threads.} Stance labels show that fear threads are argued over rather than one-sided. Replies under fear videos are about three times as likely to be corrective (13.9\% vs.\ 4.4\%, $h=0.34$) and nine times as likely to amplify alarm (9.1\% vs.\ 1.0\%, $h=0.41$), and replies to positive parents are more often disagreements (17.8\% vs.\ 6.8\%). Both gaps hold for every parent sentiment and in video-clustered regressions. Corrective replies receive similar numbers of likes in fear and non-fear threads (Cliff's $\delta=-0.05$, n.s.). For example, to a commenter who wrote that US food ``has its issues but none of these are it,'' one reply added ``We have legit issues with out food processing but what OP said is just wrong'' (labeled corrective), while in another fear thread a reply escalated a cancer claim: ``I also work at a well known chicken plant and I can concur. Chicken have cancer, the bird flu, and cyst'' (labeled amplifying).
In our 100-reply human annotation, about four in five replies that Gemini flagged as corrective or amplifying were confirmed (precision 0.81 and 0.82), so the gaps are too large to be explained by label errors.

\subsection{Fear Discourse Targets Industrial Food and Institutions (RQ3)}

\begin{figure}[t]
\centering
\includegraphics[width=0.9\columnwidth]{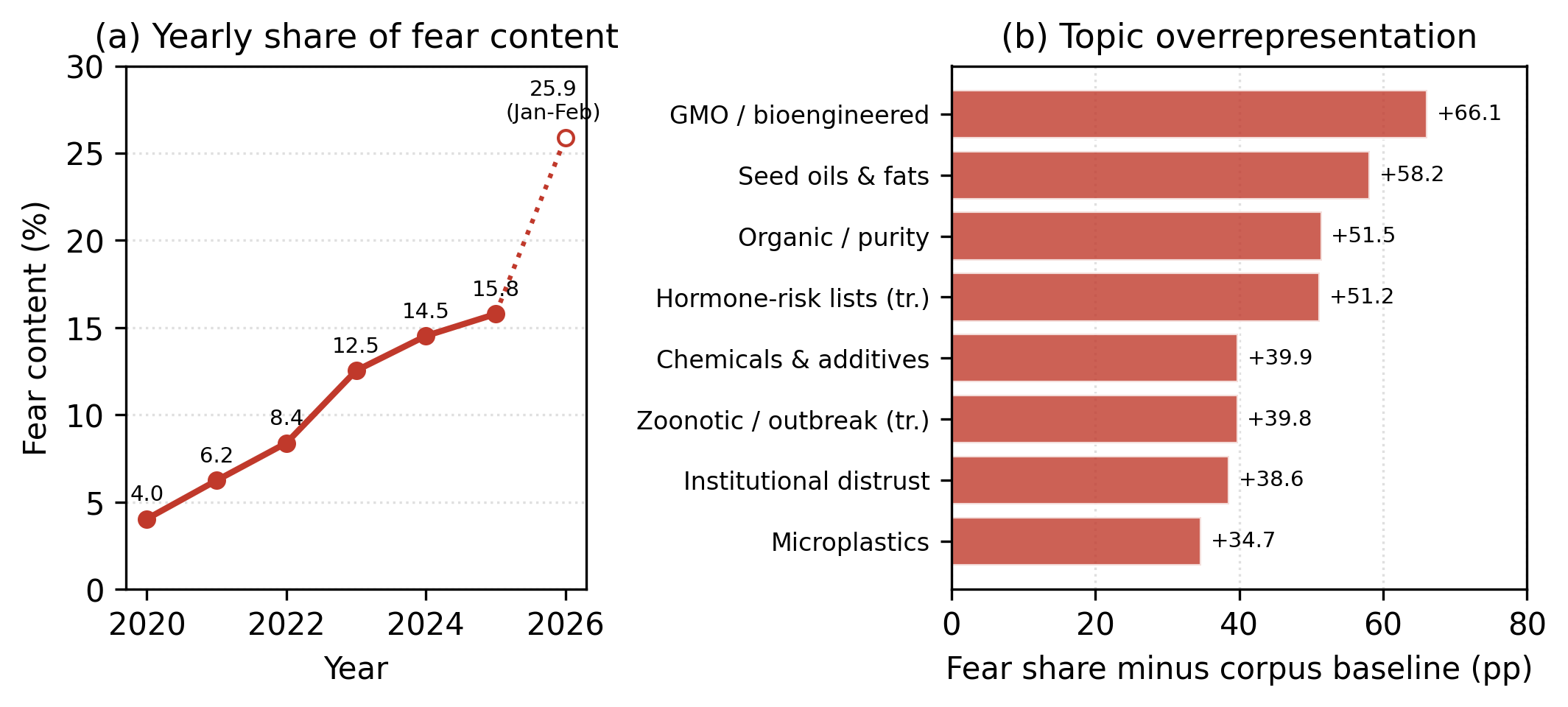}
\caption{What fear discourse is about. (a) Yearly share of fear content in the
corpus (2026 covers January--February only) (b) Fear share of the most overrepresented topic categories minus the corpus baseline (18.3\% for comments, 14.7\% for transcripts, marked ``tr.'').}
\label{fig:rq3}
\end{figure}

\textbf{What fear discourse is about.} Fear content concentrates in topics about industrial ingredients, contaminants, and distrust (Table~\ref{tab:c4}; Figure~\ref{fig:rq3}b). 18.3\% of comments come from fear videos, GMO and ``bioengineered'' ingredients (84\%), seed oils (76\%), and organic/purity skepticism (70\%) are the most fear-dominated categories, followed by chemicals and additives, institutional distrust (FDA, ``Big Pharma,'' depopulation claims), microplastics, ultra-processed food, and detox and parasite cleanses. Among transcripts, hormone-risk food lists and zoonotic-outbreak framing are similarly skewed. These categories are stable across topic-model random seeds (Section~\ref{sec:robust}). Comments in these topics read like this: ``i feel so betrayed by the fda why are they allowing this they're supposed to protect us'' and ``America likes to keep it people sick if we aren't sick there's no money in big Pharma'' (both institutional distrust); ``For the love of God people\ldots{}read all the labels\ldots{}EVERYTHING has bioengineered ingredients'' (GMO and bioengineered). The same topics also hold pushback, such as ``Ffs, seed oils are not the problem.'' We treat these categories as descriptive evidence about content, not as causal claims.

\begin{table}[t]
\centering
\small
\caption{Topic categories in which fear content is overrepresented. Lift is
percentage points over the corpus baseline (comments 18.3\%, transcripts 14.7\%).
Comment categories group 1--7 topics each.}
\resizebox{\columnwidth}{!}{%
\begin{tabular}{llrrr}
\toprule
Mode & Category & Docs & Fear \% & Lift \\
\midrule
comments & GMO / bioengineered ingredients & 598 & 84.4 & +66.1 \\
comments & Seed oils \& industrial fats & 837 & 76.5 & +58.2 \\
comments & Organic / purity skepticism & 387 & 69.8 & +51.5 \\
comments & Chemicals, pesticides, additives & 2{,}885 & 58.2 & +39.9 \\
comments & Institutional distrust & 2{,}520 & 56.9 & +38.6 \\
comments & Microplastics \& contaminants & 445 & 53.0 & +34.7 \\
comments & Processed \& fast-food avoidance & 1{,}156 & 52.0 & +33.7 \\
comments & Toxicity \& food-safety risk & 1{,}722 & 51.9 & +33.6 \\
comments & Metabolic \& chronic-health concern & 2{,}689 & 48.7 & +30.4 \\
comments & Detox, cleanses, parasites & 1{,}003 & 47.3 & +29.0 \\
comments & Body metrics \& health anxiety & 631 & 46.6 & +28.3 \\
\midrule
transcripts & Hormone-risk food lists & 91 & 65.9 & +51.2 \\
transcripts & Zoonotic / outbreak risk & 33 & 54.5 & +39.8 \\
\bottomrule
\end{tabular}
}
\label{tab:c4}
\end{table}

\begin{figure*}[t]
\centering
\includegraphics[width=0.9\textwidth]{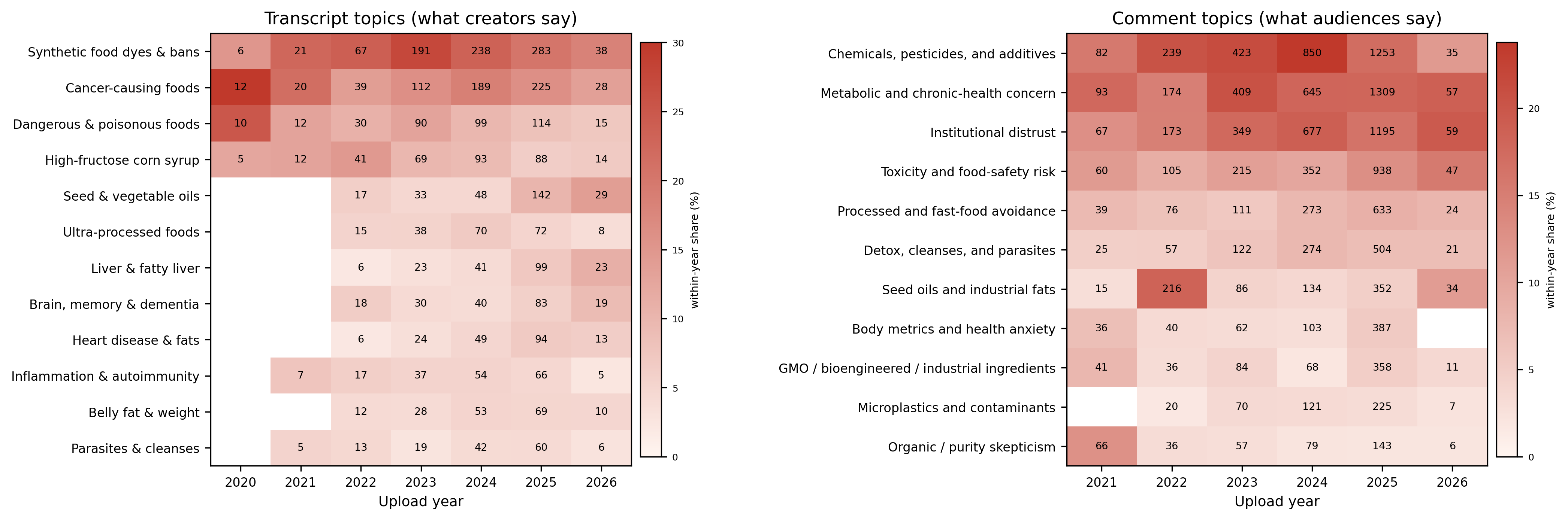}
\caption{Topic evolution of fear discourse by year for creator (transcript, left)
and audience (comment, right) topics. Cell annotation is the per-year document
count; color is the within-year share among the shown fear topics (each panel
scaled independently); cells with fewer than 5 documents are blank. The comment
panel omits 2020, which has fewer than 50 categorized comments.}
\label{fig:evolution}
\end{figure*}

\textbf{How fear discourse changes.} The dominant frames shift over time as shown in Figure~\ref{fig:evolution}. Creators begin with cancer-risk and dangerous-food framing (2020--21) and broaden into a dense set of specific scare frames, including food dyes ``banned in Europe,'' high-fructose corn syrup, seed oils, trans fats, and inflammation, that grow through 2023--2025. The discourse also \emph{diversifies}: the number of distinct fear-overrepresented transcript topics active per year grows from 5--7 in 2020--21 to the low twenties from 2022 onward, a widening repertoire of scares rather than the repetition of a few. Audiences move in parallel: organic and GMO purity talk is most prominent early (2021), seed-oil discussion spikes in 2022, and chemicals and additives, institutional distrust, and detox and microplastics talk grow through 2024--25. The share of fear content in the corpus also rises each year (4.0\% in 2020 to 15.8\% in 2025), but because keyword composition and API coverage vary across years, we read this trend as suggestive only.

\subsection{Robustness Checks}
\label{sec:robust}
\begin{table}[t]
\centering
\footnotesize
\caption{Robustness checks.}
\begin{tabular}{p{0.42\columnwidth}p{0.48\columnwidth}}
\toprule
Check & Result \\
\midrule
Controls (views, duration, year, keyword group) & All three headline gaps hold (share 1.71$\times$, $t_{50}$ 1.62$\times$, negativity +9.6 pp) \\
Within each keyword set & All hold; share rate (PPML) borderline in control set ($p=0.09$, 739 fear videos) \\
Within creator (creator FE) & Share 1.22$\times$ ($p<10^{-4}$); $t_{50}$ 1.07$\times$ and negativity +3.0 pp n.s.\ ($n=634$, 438) \\
Right-censoring (videos $\geq$1 year old) & $t_{50}$ 1.68$\times$ adjusted \\
Collection cap ($<$500 comments) & $t_{50}$ 1.62$\times$ adjusted \\
Near-complete threads ($\geq$90\% coverage) & $t_{50}$ 1.50$\times$ adjusted ($n=590$ fear) \\
Prompt without keyword information (770 videos) & Labels agree on 98.4\% ($\kappa=0.94$); Fear precision 0.72 with vs.\ 0.71 without \\
$t_{50}$ by min.\ comments, year, view bin & 6/6, 5/7, 3/4 cuts positive and significant \\
Stricter labels (LLM confidence $=1.0$, Fear precision 0.94) & Gaps grow: share 2.99$\times$, $t_{50}$ 2.17$\times$, negativity +15.4 pp \\
Human labels (817 videos) & Share 2.68$\times$, negativity +5.6 pp (both sig.); $t_{50}$ 1.23$\times$ n.s.\ ($n=96$ fear) \\
Sentiment errors (208 human labels) & Negativity gap after correcting for model errors: +11.5 pp [1.9, 19.6] vs.\ +11.6 pp raw \\
Stance errors (100 human labels) & Stance $\kappa=0.81$; corrective / amplifying precision 0.81 / 0.82 vs.\ 3$\times$ / 9$\times$ gaps \\
Topic seeds & Full-corpus refits: ARI 0.77--0.80 vs.\ main run; $\geq$96\% of each category's comments stay in fear-overrepresented topics (10/11 categories) \\
\bottomrule
\end{tabular}

\label{tab:robust}
\end{table}

Table~\ref{tab:robust} provides the results from our robustness check. The most important check compares \textit{Fear} and \textit{Non-Fear} videos posted by the same creator. Only 1{,}620 creators posted both, partly because our sampling capped each creator at two videos to keep any single account from dominating the corpus, and we found that the estimates shrink: the share-rate gap remains significant (1.22$\times$), while the half-life and negativity gaps keep their sign but lose significance. Part of the video-level differences may therefore reflect which creators produce fear content, a creator-level pattern we cannot separate further because we did not collect creator follower counts (the Research API video endpoint we used does not return them; its separate user-information endpoint does).
Additionally, right-censoring works against our finding, because fear videos are more often under a year old at collection (37.6\% vs.\ 28.1\%); restricting to fully observed videos leaves the half-life gap essentially unchanged (1.68$\times$). Incomplete comment collection does not explain the gap either: among videos whose collected comments cover at least 90\% of the reported count, fear threads still last 1.50$\times$ longer after controls (1.68$\times$ when also restricted to videos at least a year old).

Furthermore, label noise works against our finding. Restricting both groups to the classifier's most confident labels raises Fear precision against human labels from 0.64 to 0.94 and makes every gap larger. Confident labels may also select the most blatant fear videos, so we read this as evidence that misclassification attenuates our estimates rather than as the size of the true effect. Using human labels alone (817 videos) reproduces the share-rate and negativity gaps; the half-life comparison has only 96 fear videos with enough comments and is inconclusive.

\textbf{Summary of findings.} Fear videos get 62\% fewer views but 1.7--2.1$\times$ more shares per view and 1.6$\times$ longer discussions, gaps that survive controls, video age, comment coverage, and classifier-error checks. Their comments are more negative (32.5\% vs.\ 20.9\%) and angrier rather than more afraid; negativity is reinforced no more than in non-fear threads (37.8\% vs.\ 37.5\%), but positivity rarely persists (39.6\% vs.\ 57.8\%), and threads are contested. The discourse targets industrial ingredients, contaminants, and institutions, with a widening set of frames.

\section{Discussion}
Our central finding is that fear content exerts influence through \emph{dynamics} rather than \emph{reach}. Fewer people see fear videos, but they are shared more per view, discussed longer, and have more negative comment sections. The shifts are modest for any single video, but they hold under controls and when we exclude videos too recent to have accumulated their full discussion, and they apply across a large and expanding category of content. The strictest within-creator comparison weakens them, so part of the pattern may lie in who produces fear content.

\textbf{Why does fear linger?} Two mechanisms could produce longer discussions, and our data cannot fully separate them. Viewers may keep returning to and arguing in fear threads, or TikTok's recommendation feed, which does not depend on followers, may keep resurfacing videos that are shared often, bringing new commenters weeks after upload. The second reading fits the share-rate gap, since sharing is a strong engagement signal, and it would make platform ranking part of why fear content persists. Either way, the persistence is real for viewers, but distinguishing the two requires exposure data that only platforms hold.

\textbf{Aligning with the virality literature.} Prior work links high-arousal emotion to sharing \citep{berger2012viral,mousavi2022effective} and negativity to consumption \citep{robertson2023negativity,zhang2024negative}. Our results concur at the level of per-view behavior, since fear content is shared more per impression, but decouple it from reach: fear videos accumulate fewer views and likes. Their influence is better described as higher per-capita engagement among a smaller, more activated audience than as broad virality.

\textbf{A fear-appeal reading.} The EPPM \citep{witte1992putting,witte2000meta} predicts defensive ``fear control'' responses (reactance, distrust) when a threat comes without efficacy. Organic food-fear videos rarely offer actionable efficacy, and their audiences respond with anger and disgust far more than with fear itself, while institutional distrust is a persistent audience theme. This pattern is consistent with fear control rather than constructive behavior change.

\textbf{Contested threads, not echo chambers.} Negativity in fear threads is not more self-reinforcing; positivity is what fails to persist. At the same time, fear threads contain more corrective replies and more amplifying ones. They are sites of argument, which matters for intervention: corrective voices are already present and receive as many likes as corrective replies in non-fear threads. Interventions therefore do not need to create counter-speech from scratch; they need to make the counter-speech that already exists more visible. 

\textbf{What platforms can do.} Soft moderation is platforms' first line of defense against problematic content: rather than removing it, they attach warning or context labels, add friction before sharing, or limit reach \citep{singhal2023sok}. Food fear fits this approach: most of it is exaggerated or misleading rather than clearly false, so soft tags suit it better than removal, and labels generally reduce belief in and sharing of misinformation \citep{mena2020cleaning,clayton2020real,martel2023misinformation}. Our results suggest three refinements. First, \emph{detection} should use persistence. Fear videos rarely trend, so systems that wait for a spike in reach will miss them, whereas a video that is shared unusually often per view, or whose comments keep arriving days after upload, is likely fear content even if its total views stay low. Because these signals accumulate from the first days after posting, they may help flag such content early. Second, \emph{friction belongs at the share button}. Sharing is the one engagement metric on which fear content wins, and interventions at the moment of sharing, such as share-time prompts or accuracy nudges, target exactly that behavior, echoing prior work \citep{singhal2023sok,pennycook2021shifting}. Interestingly, TikTok COVID-19 labels were applied largely by hashtag, which mislabeled many videos \citep{ling2023learn}; food fear has no fixed hashtag, so labels would need to key on the claims themselves, such as the hooks that we provide in Table~\ref{tab:hooks}. Third, \emph{labels must be applied consistently}: labeling only some false claims can make unlabeled ones seem more credible \citep{pennycook2020implied}, and the engagement effects of labels are mixed \citep{zannettou2021won,papakyriakopoulos2022impact}. Because the discourse keeps adding new frames, prebunking that teaches viewers to recognize the manipulation techniques themselves, such as the promise of hidden truth or the ``banned in other countries'' comparison, may generalize better than chasing each new claim \citep{roozenbeek2022psychological}.

\textbf{How communities can help.} Fear threads already contain the replies that corrective measures can use. We found that corrective replies are three times as common as in non-fear threads. Crowds of non-professionals can rate accuracy about as reliably as fact-checkers \citep{allen2021scaling}, and posts that community notes mark as misleading are reshared less \citep{drolsbach2023diffusion,wojcik2022birdwatch}. Community notes have struggled because they arrive too late for content that spreads fast \citep{chuai2024community}; food fear is the opposite case, since its discussion lasts for days to weeks, which leaves time for crowd-sourced context to matter. Corrections from users and from experts both reduce health misperceptions \citep{bode2018see,vraga2017using}, and expert corrections are the most effective \citep{walter2021evaluating}. On TikTok, health professionals already debunk trending claims by stitching or duetting the original video \citep{sharevski2024debunk}, and such debunking videos reduce belief in the false claim \citep{bhargava2023effective}.
Platforms can amplify this work. Because the first replies to a post shape the tone of the whole thread \citep{aleksandric2022power,salehabadi2022user}, and because positive comments in fear threads rarely stay positive, surfacing well-received corrective or expert replies near the top of a fear thread could counter the damping of positivity we observe. 

\section{Limitations and Future Work}
Our corpus is keyword-defined, so it is not a random sample of TikTok food content, and its composition varies across years. The Fear labels come from an LLM classifier with moderate precision, which tends to attenuate group differences. Sentiment, emotion, and stance come from automatic classifiers trained or prompted outside this domain and may carry biases. Our sentiment and stance checks each use a single annotator (208 comments and 100 replies), and the emotion labels are not human-validated. We do not observe creator follower counts, and the within-creator comparison is underpowered. Comment collection is incomplete (median 67\% coverage) and capped. We observe when comments arrive but not when or to whom TikTok's recommendation feed showed a video, so we cannot separate audiences returning to a video from the platform resurfacing it to new viewers. 

\textbf{Future Work.} We plan to test interventions directly, for example whether context labels or highlighted expert replies change how fear threads unfold, and to follow individual creators and audiences over time to separate creator effects from content effects. We also plan to extend the analysis to other short-video platforms such as \textit{YouTube Shorts} and \textit{Instagram Reels}, and to other languages.

\textbf{Ethics.} We use publicly available content: video metadata from TikTok's Research API, and public video files, comments, and replies. We analyze content in aggregate, do not profile individual users, and report no personally identifying information. We will release video IDs, labels, the codebook, the prompt, and analysis code, not raw comment text.

\section{Conclusion}
Food-fear content on TikTok does not spike; it lingers. Across 97{,}234 videos and more than a million comments, fear videos get 62\% fewer views than other food videos, yet viewers share them 1.7--2.1$\times$ more per view and keep discussing them 1.6$\times$ longer. Their comment sections are angrier and more disgusted, and positive comments rarely stay positive (39.6\% vs.\ 57.8\%), but they are not echo chambers: negativity is reinforced no more than in non-fear threads, and corrective replies are three times as common. The fears center on industrial ingredients, contaminants, and the institutions that regulate food (``i feel so betrayed by the fda''), and the number of distinct fear frames has grown from a handful in 2020 to about twenty a year. 

These results matter because they change where platforms and researchers should look for harmful health content. Detection systems and fact-checking pipelines are tuned to content that trends; food fear rarely does, so it passes beneath them while still shaping how people talk about what they eat for weeks. Our measures of persistence, such as per-view sharing and the slow decay of discussion, offer a way to find this content early. Our reply-level results show that these threads are not closed echo chambers but contested spaces where corrective voices already exist, which makes them a promising place for soft moderation and community-based correction. 
Because nearly everyone makes food choices every day, the slow spread of food fear deserves the same attention that fast-moving misinformation already receives.

\begin{acks}
This paper is based upon work supported by the Khoury Teaching Faculty Research Seed Grant Program.
\end{acks}

\bibliographystyle{ACM-Reference-Format}
\bibliography{references}

\appendix
\section{Codebook}
\label{app:codebook}
Annotators watched each video and labeled it Fear or Non-Fear with a confidence
score (1--5). \textbf{Fear}: content that uses fear-inducing language, alarmist
claims, urgency, or sensationalized framing to create unnecessary fear about food
or nutrition. Indicators:
\begin{enumerate}
\item Alarmist terminology (``toxic,'' ``poison,'' ``deadly,'' ``killing you'').
\item Urgency (``stop eating NOW,'' ``before it's too late'').
\item Absolutist claims (``never eat this,'' ``the worst thing you can eat'').
\item Hidden-truth framing (``what they don't tell you,'' ``exposed'').
\item Conspiracy undertones (``banned in other countries,'' ``FDA is lying'').
\item Alarming on-screen text (all-caps warnings, red overlays).
\item Fear imagery (skull or hazard icons, red warning symbols).
\item Dramatic editing (sudden zooms, flashing, shaking).
\item Shock visuals (disturbing before/after images).
\item Rejection imagery (products crossed out or thrown away).
\end{enumerate}
\textbf{Decision rule}: two or more indicators $\rightarrow$ Fear; one indicator
$\rightarrow$ Fear if the overall tone is alarming, otherwise Non-Fear; none
$\rightarrow$ Non-Fear. Guiding question: ``Would a reasonable person feel
unnecessarily fearful about food after watching this?'' \textbf{Non-Fear} also
covers responsibly presented recalls and allergen alerts, educational and recipe
content, balanced health information, and satire or debunking of fear claims.
Videos that start educational but end with fear tactics are Fear; videos that
claim to debunk but reinforce fear are Fear.

\section{Analysis Samples}
\begin{table}[H]
\centering
\small
\caption{Dataset and analysis sample sizes.}
\begin{tabular}{lr}
\toprule
Dataset / analysis & Sample size \\
\midrule
Classified videos & 97{,}234 \\
\quad Fear & 12{,}698 \\
\quad Non-Fear & 84{,}536 \\
Comments in the 0--365 day window & 1{,}146{,}987 \\
Sentiment-labeled comments & 728{,}560 \\
Reply--parent sentiment pairs & 271{,}960 \\
Emotion-labeled comments (sample) & 150{,}000 \\
Stance-labeled replies (sample) & 3{,}000 \\
Transcript BERTopic documents & 76{,}713 \\
Comment BERTopic documents & 649{,}551 \\
\bottomrule
\end{tabular}
\label{tab:data}
\end{table}

\section{Keywords}
\label{app:keywords}
\noindent\textbf{Fear (17):} artificial ingredients; avoid these foods; cancer food;
contaminated food; dangerous food; food additives; food poison; foods to avoid;
harmful food; hidden ingredients; high fructose corn syrup; never eat this; poison
food; stop eating; toxic food; what happens when you eat; worst foods.

\noindent\textbf{Control (22):} balanced diet; batch cooking; cooking tips; easy
healthy meals; fiber rich foods; food education; healthy breakfast ideas; healthy
food swaps; healthy lifestyle tips; healthy lunch ideas; healthy recipes; healthy
snacks; how to eat healthy; kitchen tips; meal prep ideas; mindful eating; nutrition
facts; plant based meals; quick healthy dinner; sustainable eating; vitamin rich
foods; weeknight dinners.

\section{Classification Prompt}
\label{app:prompt}
We used the following prompt for classifying fear vs. non-fear videos. 
\begin{lstlisting}
Watch this TikTok video carefully and classify it as either **Fear** or **Non-Fear** based on BOTH what you hear AND what you see.

**Fear**: Content that uses fear-inducing language, alarmist claims, urgency, or sensationalized framing to create unnecessary fear about food or nutrition.

Must have 2 or more of these indicators:
1. Alarmist terminology: "toxic", "poison", "deadly", "killing you", "destroying your health"
2. Urgency language: "stop eating NOW", "before it's too late", "throw this away immediately"
3. Absolutist claims: "never eat this", "always avoid", "the worst thing you can eat"
4. Hidden-truth framing: "what they don't tell you", "the secret they're hiding", "exposed"
5. Conspiracy undertones: "banned in other countries", "FDA is lying", "big food doesn't want you to know"
6. Alarming on-screen text: ALL CAPS warnings, exclamation marks, red text overlays
7. Fear imagery: skull/danger icons, red warning symbols, hazard imagery
8. Dramatic editing: sudden zoom-ins, flashing effects, shaking screen
9. Shock visuals: disturbing before/after images, fear-inducing visuals
10. Rejection imagery: products shown with X marks, crossed out, thrown in trash

**Non-Fear**: Everything else - neutral, balanced, educational, recipe content, or legitimate warnings (FDA recalls, allergen alerts) presented responsibly.

Key question: "Would a reasonable person feel unnecessarily fearful about food after watching this?"

**Video Metadata:**
- Search keyword: {keyword}
- Keyword group: {keyword_group}

**Response Format (JSON only, no markdown):**
{"label": "Fear" or "Non-Fear", "confidence": 0.5-1.0, "indicators": [...], "reasoning": "brief explanation (1-2 sentences)"}
\end{lstlisting}

\section{Full Regression Results}

Table~\ref{reg:text} shows the results of our full regression test. 

\begin{table}[H]
\centering
\scriptsize
\caption{Fear coefficients. Controls: log views, duration bin, upload-year and
keyword-group fixed effects; $t_{50}$ and negativity also control for log comment
count. Standard errors clustered by creator. Within creator adds creator fixed
effects for the 1{,}620 creators with both Fear and Non-Fear videos.}
\begin{tabular}{llrl}
\toprule
Outcome & Specification & Fear effect & 95\% CI \\
\midrule
Share rate (PPML) & Unadjusted & 2.44$\times$ & [1.99, 3.00] \\
 & Controls & 2.05$\times$ & [1.67, 2.51] \\
 & Fear keywords only & 2.15$\times$ & [1.75, 2.64] \\
 & Control keywords only & 1.31$\times$ & [0.96, 1.79] \\
 & Within creator & 1.31$\times$ & [0.95, 1.79] \\
Share rate (log OLS) & Unadjusted & 2.34$\times$ & [2.28, 2.41] \\
 & Controls & 1.71$\times$ & [1.67, 1.76] \\
 & Fear keywords only & 1.70$\times$ & [1.65, 1.74] \\
 & Control keywords only & 1.58$\times$ & [1.43, 1.74] \\
 & Within creator & 1.22$\times$ & [1.11, 1.34] \\
Comment $t_{50}$ (log OLS) & Unadjusted & 1.53$\times$ & [1.42, 1.66] \\
 & Controls & 1.62$\times$ & [1.50, 1.74] \\
 & Fear keywords only & 1.57$\times$ & [1.45, 1.70] \\
 & Control keywords only & 2.06$\times$ & [1.59, 2.65] \\
 & Within creator & 1.07$\times$ & [0.72, 1.58] \\
Negative-comment rate (OLS) & Unadjusted & +12.9 pp & [12.1, 13.7] \\
 & Controls & +9.6 pp & [8.7, 10.4] \\
 & Fear keywords only & +9.2 pp & [8.3, 10.1] \\
 & Control keywords only & +12.9 pp & [10.2, 15.7] \\
 & Within creator & +3.0 pp & [$-$1.0, 7.1] \\
\midrule
Emotion: anger (logit) & Controls & OR 1.49 & [1.37, 1.61] \\
Emotion: disgust (logit) & Controls & OR 1.49 & [1.38, 1.60] \\
Emotion: fear (logit) & Controls & OR 1.21 & [1.09, 1.34] \\
\bottomrule
\end{tabular}
\label{reg:text}
\end{table}

\section{Comment Topic Categories}
\label{app:topics}

Table~\ref{tab:topics-appendix} shows our comment topic categorization. Each row groups the BERTopic topics (no automatic merging) that we labeled by reading sampled comments; ``Topics'' lists those labels. Comments column shows the number of comments in the category, and the Fear \% column shows the share that comes from fear videos (corpus baseline 18.3\%). 

\begin{table*}[t]
\centering
\caption{Comment topic categories}
\label{tab:topics-appendix}
\small
\begin{tabular}{p{0.2\textwidth}p{0.56\textwidth}rr}
\toprule
Category & Topics (labels from reading sampled comments) & Comments & Fear \% \\
\midrule
GMO / bioengineered & GMO crops debate; ``bioengineered'' label alarm & 598 & 84.4 \\
Seed oils \& fats & canola/rapeseed; seed oils harmful?; palm oil; vegetable oil is bad & 837 & 76.5 \\
Organic / purity & organic vs.\ pesticides & 387 & 69.8 \\
Chemicals \& additives & chemicals in food; Red 40 and dyes; banning dyes; US vs.\ Europe banned ingredients; US vs.\ EU Fanta; ingredient-scanner apps & 2{,}885 & 58.2 \\
Institutional distrust & FDA approves what others ban; Big Pharma; Gates/WEF; depopulation; sickness is profitable; never trust the government; ``they are trying to kill us'' & 2{,}520 & 56.9 \\
Microplastics \& contaminants & microplastics; lead in dishes; radioactive seafood & 445 & 53.0 \\
Processed \& fast food & ultra-processed food; avoiding fast food & 1{,}156 & 52.0 \\
Toxicity \& food safety & raw-food travel hygiene; ceviche/raw shrimp illness; food is poison; worms in food & 1{,}722 & 51.9 \\
Metabolic \& chronic health & autoimmune; eczema/psoriasis; inflammation; cancer; fatty liver; kidney disease & 2{,}689 & 48.7 \\
Detox \& parasites & parasite cleanses; detox/heavy-metal binders; activated charcoal & 1{,}003 & 47.3 \\
Body metrics & weight/height adequacy & 631 & 46.6 \\
\bottomrule
\end{tabular}
\end{table*}

\end{document}